\documentclass[aps,prd,superscriptaddress,showkeys,showpacs,twocolumn,a4paper]{revtex4-1}
\usepackage{ulem}
\usepackage{color}
\usepackage{graphicx}
\usepackage{hyperref}   
\usepackage{appendix}
\usepackage{epsfig}
\usepackage{epstopdf}
\usepackage{amsmath}
\usepackage{subfig}
\usepackage{comment}
\usepackage[dvipsnames]{xcolor}
\newcommand{\be}{\begin{equation}}
\newcommand{\ee}{\end{equation}}
\newcommand{\ba}{\begin{eqnarray}}
\newcommand{\ea}{\end{eqnarray}}

\begin{document}

\title{A study of charm quark dynamics in quark-gluon plasma with $3+1$D viscous hydrodynamics}

\author{Manu Kurian}
\email{manu.kurian@iitgn.ac.in}
\affiliation{Indian Institute of Technology Gandhinagar, Gandhinagar-382355, Gujarat, India}
\affiliation{Department of Physics, McGill University, 3600 University Street, Montreal, QC, H3A 2T8, Canada}

\author{ Mayank Singh}
\email{mayank.singh@mail.mcgill.ca}
\affiliation{Department of Physics, McGill University, 3600 University Street, Montreal, QC, H3A 2T8, Canada}

\author{Vinod Chandra}
%\email{vchandra@iitgn.ac.in}
\affiliation{Indian Institute of Technology Gandhinagar, Gandhinagar-382355, Gujarat, India}

\author{Sangyong Jeon}
%\email{jeon@physics.mcgill.ca }
\affiliation{Department of Physics, McGill University, 3600 University Street, Montreal, QC, H3A 2T8, Canada}

\author{Charles Gale}
%\email{gale@physics.mcgill.ca }
\affiliation{Department of Physics, McGill University, 3600 University Street, Montreal, QC, H3A 2T8, Canada}

%%%%%%%%%%%%%%%%%%%%%%%%%%%%%%%%%%%%%%%%%%%%%%%%%%%%%%%%%%%%%%%%%%%%%

\begin{abstract}
The drag and diffusion coefficients are studied within the framework of Fokker-Planck dynamics for the case of a charm quark propagating in an expanding quark-gluon plasma. The space-time evolution of the nuclear matter created in the relativistic heavy-ion collision is modelled using MUSIC, a $3+1$D relativistic viscous hydrodynamic approach. The effect of viscous corrections to the heavy quark transport coefficients is explored by considering scattering processes with thermal quarks and gluons in the medium. It is observed that the momentum diffusion of the heavy quarks is sensitive to the shear and bulk viscosity to entropy ratios. The collisional energy loss of the charm quark in the viscous quark-gluon plasma is analyzed.
\end{abstract}

%%%%%%%%%%%%%%%%%%%%%%%%%%%%%%%%%%%%%%%%%%%%%%%%%%%%%%%%%%%%%%%%%%%%


\maketitle
%%%%%%%%%%%%%%%%%%%%%%%%%%%%%%%%%%%%%%%%%%%%%%%%%%%%%%%%%%%%%%%%%%%%
\section{Introduction}
%%%%%%%%%%%%%%%%%%%%%%%%%%%%%%%%%%%%%%%%%%%%%%%%%%%%%%%%%%%%%%%%%%%%
The heavy-ion collision experiments pursued at the Relativistic Heavy Ion Collider (RHIC) at Brookhaven National Laboratory and at the Large Hadron Collider (LHC) at CERN have confirmed the existence of a new state of matter: the Quark-Gluon Plasma (QGP)~\cite{STAR,Aamodt:2010pb}. The success of hydrodynamics in describing the space-time evolution of the QGP opened new horizons in the study of relativistic heavy-ion collisions~\cite{Gale:2013da}. Early works focused on ideal hydrodynamics~\cite{Heinz}, and later the dissipative effects in the QGP evolution were incorporated and helped to explain the quantitative behaviour of experimental observables in the heavy-ion collisions~\cite{Teaney:2003kp, Romatschke:2007mq}. Several studies have been done in the determination of shear viscosity to entropy ratio $\eta/s$ from the final hadron data. Recently, the significance of non-zero bulk viscosity to entropy ratio $\zeta/s$ in the evolution of the QGP has also been emphasized~\cite{Ryu:2015vwa}.

Heavy quarks (HQs), namely charm and bottom, serve as effective probes to investigate the properties of the QGP~\cite{Prino:2016cni, Aarts:2016hap,Andronic:2015wma}, as they are mostly created in the initial moments of the collision via hard scattering. The thermalization time of HQs is estimated in the order of $10-15$ fm/c for the charm and $25-30$ fm/c for bottom quarks created at the RHIC and the LHC~\cite{Moore:2004tg,vanHees:2005wb,Cao:2011et}. This means that the HQs can report on the QGP evolution, as the lifetime of the QGP is expected in the order of $4-5$ fm/c at the RHIC~\cite{Heinz:2002gs} and about $10-12$ fm/c at the LHC~\cite{Foka:2016vta}. The HQs are propagating through the QGP while interacting with the constituent particles and can be treated with Boltzmann transport. Because of their large mass as compared to the QGP temperature scale, the scattering of HQs is amenable to a treatment in terms of Brownian motion~\cite{Das:2013kea,Li:2019wri}. The relativistic Boltzmann equation reduces to the Fokker-Planck equation under the constraint of soft momentum transfer in the HQ-thermal particle interactions and has been used to describe the propagation of HQ in the QGP~\cite{Svetitsky:1987gq,Rapp:2009my,Mrowczynski:2017kso,Song:2019cqz}. The interactions of the HQs with other quarks and gluons can be incorporated in the drag and diffusion coefficients. The HQ drag force can be related to the collisional energy loss in the medium in the formulation of the Fokker-Planck equation~\cite{Mustafa:2004dr}. There have been several attempts to study the dynamics of HQs within the scope of Brownian motion and to interpret related physical observables such as nuclear suppression factor $R_{AA}$, heavy baryon to meson ratio and elliptic flow~\cite{vanHees:2007me,Gossiaux:2008jv,Das:2009vy, Alberico:2013bza, Young:2011ug,Cao:2013ita, Cao:2013ita,Kurian:2020kct,Das:2015ana,Singh:2018wps,Adare:2006nq,Cao:2018ews}. However, many calculations supposed the QGP is a static and thermalized medium. In Ref.~\cite{GolamMustafa:1997id}, propagation of the charm quark in the equilibrating medium is investigated by considering a purely longitudinal boost-invariant expansion of the system. Recently, the radiative  energy loss of the HQ is further studied in the longitudinal expansion~\cite{Sarkar:2018erq}. It is therefore an interesting task to investigate the HQ dynamics with a realistic description of the viscous QGP evolution.

The focus of the current analysis is to investigate the HQ drag and momentum diffusion in the expanding viscous QGP, and explore the sensitivity of HQ transport coefficients and collisional energy loss to a non-zero viscosity to entropy ratio. This requires relativistic hydrodynamical modelling of the evolution of the medium created in the relativistic heavy-ion collision. The viscous hydrodynamic equations up to second order in flow velocity gradients are the standard input to characterize the bulk medium created in the collisions~\cite{Baier:2007ix,Betz:2009zz,Florkowski:2015lra}. This investigation incorporates the viscous effects in the HQ dynamics in the QGP that enters through the momentum distribution of constituent particles in the medium and through the screening mechanism. A collision integral that takes account of the $2\rightarrow 2$ elastic HQ-thermal particle collisions in the QGP medium is considered in the analysis. The significance of viscous coefficients of the QGP medium has already been discussed in dilepton emission, photon production, heavy quarkonia, anisotropic flow and other relevant observables of heavy-ion collisions at the RHIC and the LHC~\cite{Paquet:2015lta,Dusling:2008xj,Bhalerao:2015iya,Schenke:2011bn,Vujanovic:2013jpa,Thakur:2020ifi,Shen:2014nfa}. 

 The rest of the article is organized as follows. In Section II, a brief description of HQ drag and momentum diffusion is presented within the framework of Fokker-Planck dynamics. Section III is devoted to the details of the relativistic hydrodynamical modeling to calculate the evolution of the background QGP, followed by the description of viscous corrections to the HQ transport coefficients. The results are discussed in Section IV and finally, we conclude in Section V.
%%%%%%%%%%%%%%%%%%%%%%%%%%%%%%%%%%%%%%%%%%%%%%%%%%%%%%%%%%%%%%%%
\section{HQ drag and diffusion}
%%%%%%%%%%%%%%%%%%%%%%%%%%%%%%%%%%%%%%%%%%%%%%%%%%%%%%%%%%%%%%%%
In the present analysis, we adopt the formalism developed by Svetitsky~\cite{Svetitsky:1987gq} to investigate the HQ dynamics in the QGP medium. 
The dynamics of HQ can be described by the relativistic Boltzmann equation as,
\begin{equation}\label{1.1}
p^{\mu}\partial_{\mu}f_{HQ}=\bigg(\dfrac{\partial f_{HQ}}{\partial t}\bigg)_c,
\end{equation}
where $f_{HQ}$ is HQ momentum distribution function. 
The term $(\frac{\partial f_{HQ}}{\partial t})_c$ denotes the collision term that quantifies the rate of change of $f_{HQ}$ due to the interactions/scattering with thermal quarks and gluons in the medium. The relativistic collision integral for the two-body collision takes the form,
\begin{align}\label{1.2}
\bigg(\dfrac{\partial f_{HQ}}{\partial t}\bigg)_c=&\int{d^3{\bf k}\bigg[\omega({\bf p}+{\bf k},{\bf k})f_{HQ}({\bf p}+{\bf k})}\nonumber\\
&-\omega({\bf p},{\bf k})f_{HQ}({\bf p})\bigg],
\end{align}
where $\omega({\bf p},{\bf k})$ is the collision rate per unit momentum phase-space of the HQ with quarks and gluons that change its momentum from ${\bf p}$ to ${\bf p}-{\bf k}$. The relativistic Boltzmann equation simplified to Fokker-Planck dynamics by employing the Landau approximation~\cite{Landau} which assumes small momentum transfer in the HQ-thermal particles scattering, 
\begin{align}\label{1.3}
\dfrac{\partial f_{HQ}}{\partial t}=\dfrac{\partial}{\partial p_i}\bigg[A_i({\bf p})~f_{HQ}
+\dfrac{\partial}{\partial p_j}\big[B_{ij}({\bf p})~f_{HQ}\big]\bigg],
\end{align}
where $A_i$ and $B_{ij}$ are the drag force and momentum diffusion of the HQs in the QGP medium. Here, $i, j=1,2,3$ denote the spatial components of the $3-$vectors. The HQ drag 
and momentum diffusion take the following forms for the process $HQ(p)+l(q)\rightarrow HQ(p^{'})+l(q^{'})$, 
where $l$ represents quarks or gluons in the medium, as
\begin{align}\label{1.4}
A_i=&\dfrac{1}{\gamma_{c}}\dfrac{1}{2P^0}\int{\dfrac{d^3{\bf q}}{(2\pi)^3 2Q^0}}\int{\dfrac{d^3 {\bf p}^{'}}{(2\pi)^32{P^{'}}^0}}\int{\dfrac{d^3{\bf q}^{'}}{(2\pi)^32{Q^{'}}^0}}\nonumber\\
&\times(2\pi)^4\delta^4(P+Q-P^{'}-Q^{'})\sum\mid\mathcal{M}_{HQ,g/q}\mid^2\nonumber\\
&\times f_{g/q}({Q})\Big(1\pm f_{g/q}({Q}^{'})\Big)\big({\bf p}-{\bf p}^{'}\big)_i\nonumber\\
& \equiv \langle\langle\big({\bf p}-{\bf p}^{'}\big)_i\rangle\rangle,
\end{align}
and
\begin{align}\label{1.5}
B_{ij}=&\dfrac{1}{2\gamma_{c}}\dfrac{1}{2P^0}\int{\dfrac{d^3{\bf q}}{(2\pi)^32Q^0}}\int{\dfrac{d^3{\bf p}^{'}}{(2\pi)^32{P^{'}}^0}}\int{\dfrac{d^3{\bf q}^{'}}{(2\pi)^32{Q^{'}}^0}}\nonumber\\
&\times(2\pi)^4\delta^4(P+Q-P^{'}-Q^{'})\sum\mid\mathcal{M}_{HQ,g/q}\mid^2\nonumber\\
&\times f_{g/q}({Q})\Big(1\pm f_{g/q}({Q}^{'})\Big)({\bf p}-{\bf p}^{'})_i ({\bf p}-{\bf p}^{'})_j\nonumber\\
&\equiv \langle\langle\dfrac{1}{2}\big({\bf p}-{\bf p}^{'}\big)_i\big({\bf p}-{\bf p}^{'}\big)_j\rangle\rangle,
\end{align}
with $\gamma_{c}$ as the statistical degeneracy of the HQ and $f_{g/q}$ is the momentum distribution of the thermal particles in the bulk medium. Here, $P=(E_p, {\bf p}), Q=(E_q, {\bf q})$ denote the energy-momenta of the HQ and thermal particles in the entrance channel and $P^{'}=(E_{p^{'}}, {\bf p}^{'}), Q=(E_{q^{'}}, {\bf q}^{'})$ represent the energy-momenta after scattering. The HQ-thermal particles $2\rightarrow 2$ scattering matrix element, $\mid\mathcal{M}_{HQ,g/q}\mid$, can be obtained from Feynman diagrams as described in Ref.~\cite{Svetitsky:1987gq}.  The drag force and momentum diffusion respectively measure the thermal average of the momentum transfer ${\bf k}={\bf p}-{\bf p}^{'}$ and its square, due to the HQ-thermal particles scattering in the QGP medium. Since $A_i$ and $B_{ij}$ depend only on ${\bf p}$, they can be decomposed as follows,
\begin{align}\label{1.6}
&A_i=p_iA(p^2, T),\\
&B_{ij}=\bigg(\delta_{ij}-\dfrac{p_ip_j}{p^2}\bigg)B_0(p^2, T)+\dfrac{p_ip_j}{p^2}B_1(p^2, T),
\end{align}
with $p^2=\mid{\bf p}\mid^2$. Here, $A$ is the HQ drag coefficient and $B_{ij}$ follows longitudinal-transverse decomposition where $B_0$ and $B_1$ denotes the independent transverse and longitudinal diffusion coefficients. The coefficients can be defined in terms of interaction amplitude as follows,
\begin{align}
&A=~\langle\langle1\rangle\rangle-{\langle\langle{\bf p}.{\bf p}^{'}\rangle\rangle}/{p^2},\label{1.610}\\
&B_{0}=\dfrac{1}{4}\Big[\langle\langle{p^{'}}^2\rangle\rangle-{\langle\langle({\bf p}.{\bf p}^{'})^2\rangle\rangle}/{p^2}\Big],\label{1.611}\\
&B_{1}=\dfrac{1}{2}\Big[{\langle\langle({\bf p}.{\bf p}^{'})^2\rangle\rangle}/{p^2}-2\langle\langle{\bf p}.{\bf p}^{'}\rangle\rangle+p^2\langle\langle1\rangle\rangle\Big].\label{1.612}
\end{align}
The integrals can be simplified by solving the kinematics in the center-of-momentum frame of the colliding particles~\cite{GolamMustafa:1997id},
\begin{align}\label{1.7}
\langle\langle&F(p^{'})\rangle\rangle=\dfrac{1}{512\pi^4\gamma_c}\dfrac{1}{E_p}\int_0^{\infty}{\dfrac{q^2}{E_q}dq}\int_{-1}^{1}{d\cos{\chi}}\nonumber\\
&\times f_{g/q}(E_q)\dfrac{\sqrt{(s+m_c^2-m^2_{g/q})^2-4sm_c^2}}{s}\int_{-1}^{1}{d\cos{\theta_{cm}}}   \nonumber\\
&\times  \sum\mid\mathcal{M}_{HQ,g/q}\mid^2\int_{0}^{2\pi}{d\phi_{cm}e^{\beta E_{q^{'}}}f_{g/q}(E_{q^{'}})F(p^{'})},
\end{align}
where $s=(E_p+E_q)^2-({\bf p}+{\bf q})^2$, $E_{q^{'}}=E_p+E_q-E_{p^{'}}$ and $p^{'}$ can be represented in terms of $p$, $q$, $\theta_{cm}$ and $\phi_{cm}$. Here, $m_c$ and $m_{g/q}$ are the mass of charm quark and thermal mass of the gluons/quarks, respectively. 

%%%%%%%%%%%%%%%%%%%%%%%%%%%%%%%%%%%%%%%%%%%%%%%%%%%%%%%%%%%%%%%%%%%%%%%%
\section{Hydrodynamical modelling and Viscous corrections to HQ dynamics}
%%%%%%%%%%%%%%%%%%%%%%%%%%%%%%%%%%%%%%%%%%%%%%%%%%%%%%%%%%%%%%%%%%%%%%%%
\subsection{Hydrodynamical evolution of the QGP}
%%%%%%%%%%%%%%%%%%%%%%%%%%%%%%%%%%%%%%%%%%%%%%%%%%%%%%%%%%%%%%%%%%%%%%%
For the purpose of this study, we consider the realistic bulk evolution history of a Pb+Pb collision event at 2.76 TeV. To illustrate the viscous effects on the charm quark dynamics, we use one event with the IP-Glasma initial state~\cite{Schenke:2012wb,McDonald:2016vlt}. The hydrodynamic phase is evolved using MUSIC, a $3+1$D hydrodynamical approach~\cite{Schenke:2010nt}.

The shear tensor $\pi^{\mu\nu}$ and bulk-viscous pressure $\Pi$ constitutes the dissipative part of the energy-momentum tensor of the QGP,
\begin{equation}
    \delta T^{\mu\nu}=\pi^{\mu\nu}-\Delta^{\mu\nu}\Pi,
\end{equation}
where $\Delta^{\mu\nu}=g^{\mu\nu}-u^{\mu}u^{\nu}$ is the projection operator orthogonal to the fluid velocity $u^{\mu}$ and $g^{\mu\nu}=\text{diag}(1, -1, -1, -1)$ is the metric tensor. It is established that the dynamics of the bulk QGP is sensitive to the viscous transport (both shear and bulk viscosity) of the medium~\cite{Bhadury:2019xdf,Vujanovic:2017psb,Schenke:2011zz}. The stress tensor and bulk-viscous pressure satisfy relaxation-type equations as follows~\cite{Denicol:2012cn,Denicol:2014vaa,
Jaiswal:2013npa}
\begin{align}
 {\tau_\pi}\dot{\pi}^{\langle\mu\nu\rangle}+\pi^{\mu\nu}=&2\eta\sigma^{\mu\nu}-\delta_{\pi\pi}\pi^{\mu\nu}\theta+\phi_{7}\pi_{\beta}^{\langle\mu}\pi^{\nu \rangle\beta} \nonumber\\
 &-\tau_{\pi\pi}\pi_{\beta}^{\langle\mu}\sigma^{\nu \rangle\beta}+\lambda_{\pi\Pi}\Pi\sigma^{\mu\nu},
\end{align}
\begin{align}
 {\tau_\Pi}\dot{\Pi}+\Pi=-\zeta\theta-\delta_{\Pi\Pi}\Pi\theta
 +\lambda_{\Pi\pi}\pi^{\mu\nu}\sigma_{\mu\nu},
\end{align}
with $\theta=\partial_{\mu}u^{\mu}$ as the expansion parameter and $\sigma^{\mu\nu}=\Delta^{\mu\nu}_{\alpha\beta}\partial^{\alpha}u^{\beta}$ where $\Delta^{\mu\nu}_{\alpha\beta}\equiv\frac{1}{2}(\Delta^\mu_\alpha\Delta^\nu_\beta +\Delta^\mu_\beta\Delta^\nu_\alpha)-\frac{1}{3}\Delta^{\mu\nu}\Delta_{\alpha\beta}$ defines the traceless, symmetric, projection operator. We  use the notation ${X}^{\langle\mu\nu\rangle}=\Delta^{\mu\nu}_{\alpha\beta}X^{\alpha\beta}$ in the viscous evolution equations. The values of shear and bulk viscosities are fixed to match the measured transverse momentum integrated anisotropic flow coefficients and the spectra of charged particles. 
The shear viscosity over entropy density is chosen as $\eta/s=0.13$. A temperature dependent bulk viscosity profile parameterized in~\cite{Denicol:2009am} and used in~\cite{Ryu:2015vwa,Paquet:2015lta} is used in the current analysis. The second-order coefficients $\delta_{\pi\pi}, \phi_{7}, \tau_{\pi\pi}, \lambda_{\pi\Pi}, \tau_{\pi}, \delta_{\Pi\Pi}, \lambda_{\Pi\pi}, \tau_{\Pi}$ are related to the first-order transport coefficients, shear and bulk viscosities, $\eta$ and $\zeta$ respectively~\cite{Denicol:2014vaa}. As the space-time evolution of the QGP is described by the viscous hydrodynamics, it is understood that the system is not exactly in thermal equilibrium. To that end, one needs to obtain the viscous corrections to the momentum distribution function of quarks and gluons while estimating the HQ transport coefficients in the viscous medium.  
%%%%%%%%%%%%%%%%%%%%%%%%%%%%%%%%%%%%%%%%%%%%%%%%%%%%%%%%%%%%%%%%%%%%
\subsection{Shear-viscous correction}
%%%%%%%%%%%%%%%%%%%%%%%%%%%%%%%%%%%%%%%%%%%%%%%%%%%%%%%%%%%%%%%%%%%%
For a given HQ-thermal particle collision process, one can include the viscous corrections to the local momentum distribution of the thermal particles and thereby to the screening Debye mass in the medium. The first step towards the estimation of the dissipative effects in the HQ evolution in the QGP is to include the viscous correction to the quark and gluons distribution function. We linearize the viscous correction in the HQ drag and momentum diffusion in the shear-stress tensor $\pi^{\mu\nu}$, yielding a leading order result in $\frac{\pi^{\mu\nu}}{\epsilon+P}$.
The distribution function takes the following form~\cite{Paquet:2015lta},
\begin{equation}\label{1.8}
f_{g/q}(Q, X)= f_{g/q}^0(Q)+\delta f_{g/q}(Q, X),   
\end{equation}
with %the non-equilibrium part defined as, 
\begin{equation}\label{1.9}
\delta f_{g/q}(Q, X)=\pi_{\mu\nu}Q^{\mu}Q^{\nu}\sum_j S^j_X(X)S_M^j(Q,T).   
\end{equation}
The Eq.~(\ref{1.9}) is the general form of the non-equilibrium part of the distribution function. Note that the sum over the index $j$ is necessary only when space and momentum dependence terms cannot be factorized directly, see the discussions in Ref.~\cite{Paquet:2015lta}. For the parton distribution function, the functions $S_X$ and $S_M$ respectively take the form,
\begin{align}
& S_X=\frac{1}{2(\epsilon+\mathcal{P})},&&S_M=\frac{f_{g/q}^0(q)(1\pm f_{g/q}^0(q))}{T^2},  
\end{align}
where $\epsilon$ and $\mathcal{P}$ are the energy density and pressure of the medium. These thermodynamical quantities are related through the equation of state (EoS) of the QGP. Linearizing in $\delta f_{g/q}$, $A_i$ in Eq.~(\ref{1.4}) that defines the thermal average of momentum transfer becomes,
\begin{align}\label{1.11}
A_{i}\simeq A_{i}^{(0)}+ A_{i}^{{\text{shear}}},   
\end{align}
in the leading order where, 
\begin{align}\label{1.12}
 A_{i}^{{\text{shear}}}&=\dfrac{1}{\gamma_{c}}\dfrac{1}{2P^0}\int{\dfrac{d^3{\bf q}}{(2\pi)^32Q^0}}\int{\dfrac{d^3 {\bf p}^{'}}{(2\pi)^32{P^{'}}^0}}\int{\dfrac{d^3{\bf q}^{'}}{(2\pi)^32{Q^{'}}^0}}\nonumber\\
&\times(2\pi)^4\delta^4(P+Q-P^{'}-Q^{'})\sum\mid\mathcal{M}_{HQ,g/q}\mid^2\nonumber\\
&\times\bigg[\delta f_{g/q}({Q})\Big(1\pm f^0_{g/q}({Q}^{'})\Big)\pm f^0_{g/q}({Q})\delta f_{g/q}({Q}^{'})\bigg]\nonumber\\
&\times\big({\bf p}-{\bf p}^{'}\big)_i.
\end{align}
The first order correction to the distribution function is described in the Eq.~(\ref{1.8}). Using Eq.~(\ref{1.9}), the effect of shear viscosity on HQ drag Eq.~(\ref{1.12}) can be written as follows,
\begin{align}\label{1.13}
 A_{i}^{{\text{shear}}}=\pi_{\mu\nu}P^{\mu}P^{\nu}\sum_j S^j_X(X)\bar{S}_M^j(P,T),
\end{align}
where $\pi^{\mu\nu}g_{\mu\nu}=\pi^{\mu\nu}u_{\mu}=0$ were employed to constrain the coefficient multiplying $\pi^{\mu\nu}$.
Following the same prescriptions as in Ref.~\cite{Paquet:2015lta}, we obtain
\begin{align}\label{1.14}
\bar{S}_M^j(P,T)&=\dfrac{1}{2[(u.P)^2-P^2]}\bigg[g_{\mu\nu}+\dfrac{P^2+2(u.P)^2}{[(u.P)^2-P^2]}u_{\mu}u_{\nu}\nonumber\\
&+3\dfrac{P^{\mu}P^{\nu}}{[(u.P)^2-P^2]}-3\dfrac{(u.P)}{[(u.P)^2-P^2]} (P^{\mu}u^{\nu}\nonumber\\
&+P^{\nu}u^{\mu})\bigg]\dfrac{1}{\gamma_{c}}\dfrac{1}{2P^0}\int{\dfrac{d^3{\bf q}}{(2\pi)^32Q^0}}\int{\dfrac{d^3 {\bf p}^{'}}{(2\pi)^32{P^{'}}^0}}\nonumber\\
&\times\int{\dfrac{d^3{\bf q}^{'}}{(2\pi)^32{Q^{'}}^0}}(2\pi)^4\delta^4(P+Q-P^{'}-Q^{'})\nonumber\\
&\times\big({\bf p}-{\bf p}^{'}\big)_i\sum\mid\mathcal{M}_{HQ,g/q}\mid^2 \bigg[Q^{\mu}Q^{\nu}S^j_M(Q, T)\nonumber\\
&\times\Big(1\pm f^0_{g/q}({ Q}^{'})\Big)\pm f^0_{g/q}(Q){Q^{'}}^{\mu}{Q^{'}}^{\nu}S^j_M(Q^{'}, T)\bigg].
\end{align}
The term $\bar{S}_M(P,T)$ is a scalar, that depends on the HQ momentum and one can evaluate this scalar in the fluid rest frame. The shear-viscous part of the HQ drag, $ A_{i}^{{\text{shear}}}$, follows the same decomposition as Eq.~(\ref{1.6}), and we can simplify the integral in the center of mass frame. Similarly, we can estimate the shear-viscous correction to the momentum diffusion of the HQ in the QGP medium using Eq.~(\ref{1.8}) in Eq.~(\ref{1.5}). To proceed further, the shear-viscous correction to the general term $\langle\langle F(p^{'})\rangle\rangle$ needs to be done.
The viscous correction to the integral in the center of mass frame can be defined as,
\begin{align}\label{1.16}
\langle\langle F(p^{'})\rangle\rangle^{\text{shear}}&=\pi_{\mu\nu}P^{\mu}P^{\nu}\dfrac{1}{(\epsilon+P)}\dfrac{1}{T^2 ~512\pi^4\gamma_c}\nonumber\\
&\times\dfrac{1}{4E_pp^2}\Big[\Gamma_1(p, T)\pm \Gamma_2(p,T)\Big],
\end{align}
where $\Gamma_1(p, T)$ and $\Gamma_2(p, T)$ take the forms,
\begin{align}\label{1.17}
\Gamma_1&=\int_0^{\infty}{\dfrac{q^2}{E_q}dq}\int_{-1}^{1}{d\cos{\chi}} \dfrac{\sqrt{(s+m_c^2-m^2_{g/q})^2-4sm_c^2}}{s}\nonumber\\
&\times f^0_{g/q}(E_q) \bigg(1\pm f^0_{g/q}(E_q)\bigg)\bigg[m^2_{g/q}+3q^2\cos^2\chi-E_q^2\bigg]\nonumber\\
&\times\int_{-1}^{1}{d\cos{\theta_{cm}}}    \sum\mid\mathcal{M}_{HQ,g/q}\mid^2 \int_{0}^{2\pi}{d\phi_{cm}}e^{\beta E_{q^{'}}}\nonumber\\
&\times f_{g/q}(E_{q^{'}})F(p^{'}),
\end{align}
and
\begin{align}\label{1.18}
\Gamma_2&=\int_0^{\infty}{\dfrac{q^2}{E_q}dq}\int_{-1}^{1}{d\cos{\chi}} \dfrac{\sqrt{(s+m_c^2-m^2_{g/q})^2-4sm_c^2}}{s}\nonumber\\
&\times f^0_{g/q}(E_q)\int_{-1}^{1}{d\cos{\theta_{cm}}}    \sum\mid\mathcal{M}_{HQ,g/q}\mid^2 \int_{0}^{2\pi}{d\phi_{cm}}\nonumber\\
&\times \bigg(1\pm f_{g/q}(E_{q^{'}})\bigg)f_{g/q}(E_{q^{'}})\bigg[m^2_{g/q}+\dfrac{3}{p^2}\big(p^2+pq\cos\chi\nonumber\\
&-({\bf p}.{\bf p}^{'})\big)^2-E_{q^{'}}^2\bigg]F(p^{'}).
\end{align}
Note that here, $p^{'}$ is a function of $p$, $q$, $\cos\chi$ and scattering angles in the center of mass frame, $\theta_{cm}$ and $\phi_{cm}$, respectively.

Viscous corrections to the distribution functions of quarks and gluons modify the gluon self-energy and hence the screening  mass in the medium. The bulk-viscous correction to the retarded gluon self energy and Debye screening mass $\mu^2\rightarrow\mu^2+\delta\mu^2$ is investigated in Ref.~\cite{Du:2016wdx}. 
The Debye mass can be defined using the gluon self-energy, 
$\mu^{2}=\Pi_{00}(q_0=0,\mid \vec{q}\mid \longrightarrow 0)$, and takes the following form
\begin{equation}\label{1.24}
\mu^{2}=4\pi\alpha_{s}\beta\int{\dfrac{d^{3}\bf{q}}{(2\pi)^{3}}\Big[2N_c f_g(1+f_g)+2N_ff_q(1-f_q)}\Big],
\end{equation}
where $\alpha_s=g^2/4\pi$ is the coupling constant, $N_f$ is the number of flavors and $N_c$ denotes the number of colors.
The viscous corrections to the screening mass can be defined from Eq.~(\ref{1.24}) as,
\begin{align}\label{1.25}
\delta\mu^{2}&=4\pi\alpha_{s}\beta\int{\dfrac{d^{3}\bf{q}}{(2\pi)^{3}}\Big[2N_c \delta f_g(1+2f^0_g)}\nonumber\\
&+2N_f\delta f_q(1-2f^0_q)\Big].
\end{align}
The first-order shear-viscous correction for the Debye screening mass can be obtained by substituting Eq.~(\ref{1.9}) in Eq.~(\ref{1.25}) and we have,
\begin{align}\label{1.27}
\delta&\mu^{2}=4\pi\alpha_{s}\pi_{\mu\nu}Q^{\mu}Q^{\nu}\dfrac{1}{(\epsilon+P)}\dfrac{1}{T^3}\int_0^{\infty}{\dfrac{q^2}{(2\pi)^2}~dq}\int_{-1}^{1}{d\cos{\chi}}\nonumber\\
&\times\Bigg[2N_c \bigg((m^2_{g}+3q^2\cos^2\chi-E_q^2)f^0_g(E_q)\Big(1+f^0_g(E_q)\Big)\nonumber\\
&\times\Big(1+2f^0_g(E_q)\Big)\bigg)+2N_f\bigg((m^2_{q}+3q^2\cos^2\chi-E_q^2)\nonumber\\
&\times f^0_q(E_q)\Big(1-f^0_q(E_q)\Big)\Big(1-2f^0_q(E_q)\Big)\bigg)\Bigg].
\end{align}
Note that $cos\chi$ appears in the integrand in only one place. Doing the $cos\chi$ integral the term vanishes, and hence we conclude that shear-viscous correction of the screening mass of the QGP is not directly affecting the HQ drag and diffusion in leading order.
%%%%%%%%%%%%%%%%%%%%%%%%%%%%%%%%%%%%%%%%%%%%%%%%%%%%%%%%%%%%%%%%%%%%
\subsection{Bulk-viscous correction}
%%%%%%%%%%%%%%%%%%%%%%%%%%%%%%%%%%%%%%%%%%%%%%%%%%%%%%%%%%%%%%%%%%%%
\subsubsection{Distribution function}
%%%%%%%%%%%%%%%%%%%%%%%%%%%%%%%%%%%%%%%%%%%%%%%%%%%%%%%%%%%%%%%%%%%%
In this section, we focus on the bulk-viscous correction to the HQ transport coefficients, considering the viscous corrections through the distribution function and screening mass in the QGP medium. For the quantitative analysis, we utilize the leading order bulk-viscous correction to the distribution function obtained from the Chapman-Enskog expansion within the relaxation-time approximation and it takes the following form, 
\begin{align}\label{2.0}
\delta f_{g/q}(Q, X)&=-\beta {f^0}_{g/q}(Q)\Big(1\pm f^0_{g/q}(Q)\Big)\bigg(E_q-\dfrac{m^2_{g/q}}{E_q}\bigg)\nonumber\\&
\times \bigg(c_s^2-\dfrac{1}{3}\bigg) \dfrac{\Pi(X)}{(\zeta/\tau_R)},
\end{align}
where 
\begin{equation}\label{2.01}
\dfrac{\zeta}{\tau_R}\approx 15\bigg(\dfrac{1}{3}-c_s^2\bigg)^2\big(\epsilon+\mathcal{P}\big),
\end{equation}
with $\tau_R$ as the thermal relaxation time and $c^2_s$ as the square of the speed of sound in the medium. It is important to note that the effect of the running of coupling is not considered in the above expression of $\delta f_{g/q}$. For the general expression of bulk-viscous correction to the distribution function while considering the running coupling and its reduction to the non-running coupling limit, see the discussion in Ref.~\cite{Paquet:2015lta}. Eq.~(\ref{2.0}) can be written as,
\begin{equation}\label{2.1}
\delta f_{g/q}(Q, X)=\Pi\sum_j B^j_X(X)B_M^j(Q,T),  
\end{equation}
 where $B_X(X)$ and $B_M(Q,T)$ for partons respectively take the forms,
\begin{align}\label{2.2}
&B_X(X)=\dfrac{1}{15\big(\frac{1}{3}-c_s^2\big)\big(\epsilon+\mathcal{P}\big)},\\
&B_M(Q,T)=\dfrac{1}{T}f^0_{g/q}(Q)\Big(1\pm f^0_{g/q}(Q)\Big)\bigg(E_q-\dfrac{m^2_{g/q}}{E_q}\bigg).
\end{align}
Employing Eq.~(\ref{2.1}) and Eq.~(\ref{1.4}), the effect of bulk viscosity on HQ drag can be defined as,
\begin{align}\label{2.4}
 A_{i}^{{\text{bulk}}}=\Pi\sum_j B^j_X(X)\bar{B}_M^j(P,T),
\end{align}
where,
\begin{align}\label{2.5}
\bar{B}_M^j&(p,T)=\dfrac{1}{\gamma_{c}}\dfrac{1}{2P^0}\int{\dfrac{d^3{\bf q}}{(2\pi)^32Q^0}}\int{\dfrac{d^3 {\bf p}^{'}}{(2\pi)^32{P^{'}}^0}}\int{\dfrac{d^3{\bf q}^{'}}{(2\pi)^32{Q^{'}}^0}}\nonumber\\
&\times(2\pi)^4\delta^4(P+Q-P^{'}-Q^{'})\big({\bf p}-{\bf p}^{'}\big)_i\sum\mid\mathcal{M}_{HQ,g/q}\mid^2 \nonumber\\
&\times\bigg[B^j_M(Q, T)\Big(1\pm f^0_{g/q}({ Q}^{'})\Big)\pm f^0_{g/q}(Q)B^j_M(Q^{'}, T)\bigg].
\end{align}
The drag coefficient $A$ can be described from Eq.~(\ref{1.610}) by following the decomposition. 
The bulk-viscous correction to the simplified integral in the center of mass frame takes the following form,
\begin{align}\label{2.6}
\langle\langle F(p^{'})\rangle\rangle^{\text{bulk}}=\dfrac{\Pi  B_X(X)}{ 512\pi^4\gamma_c}\dfrac{1}{E_p}\Big[\Lambda_1(p, T)\pm \Lambda_2(p,T)\Big],
\end{align}
where,
\begin{align}\label{2.7}
\Lambda_1&=\int_0^{\infty}{\dfrac{q^2}{E_q}dq}\int_{-1}^{1}{d\cos{\chi}} \dfrac{\sqrt{(s+m_c^2-m^2_{g/q})^2-4sm_c^2}}{s} \nonumber\\
&\times B_M(Q, T)\int_{-1}^{1}{d\cos{\theta_{cm}}}\sum\mid\mathcal{M}_{HQ,g/q}\mid^2 \int_{0}^{2\pi}{d\phi_{cm}}\nonumber\\
&\times e^{\beta E_{q^{'}}}f_{g/q}(E_{q^{'}})F(p^{'}),
\end{align}
and
\begin{align}\label{2.8}
\Lambda_2&=\int_0^{\infty}{\dfrac{q^2}{E_q}dq}\int_{-1}^{1}{d\cos{\chi}} \dfrac{\sqrt{(s+m_c^2-m^2_{g/q})^2-4sm_c^2}}{s} \nonumber\\
&\times f^0_{g/q}(E_q)\int_{-1}^{1}{d\cos{\theta_{cm}}}    \sum\mid\mathcal{M}_{HQ,g/q}\mid^2 \int_{0}^{2\pi}{d\phi_{cm}}\nonumber\\
&\times B_M(Q^{'}, T)F(p^{'}).
\end{align}
%%%%%%%%%%%%%%%%%%%%%%%%%%%%%%%%%%%%%%%%%%%%%%%%%%%%%%%%%%%%%%%%%%%%
\subsubsection{Bulk-viscous correction to screening mass}
%%%%%%%%%%%%%%%%%%%%%%%%%%%%%%%%%%%%%%%%%%%%%%%%%%%%%%%%%%%%%%%%%%%%
The bulk-viscous correction to the distribution function of quarks and gluons in the QGP medium modifies the Debye screening mass~\cite{Du:2016wdx}, which in turn affects the collision matrix element for the $2\rightarrow 2$ HQ-gluon/quark scattering process. For the t-channel HQ-quark or antiquark scattering process, the matrix element takes the following form~\cite{Svetitsky:1987gq},
\begin{equation}\label{2.9}
\mid\mathcal{M}_{HQ,q}\mid^2=256N_f\pi^2\alpha_s^2\dfrac{(m_c^2-s)^2+(m_c^2-u)^2+2m_c^2t}{(t-\mu^2)^2},    
\end{equation}
where $s, u, t$ are Mandelstam variables. Incorporating the effect of leading order bulk-viscous correction to the Debye screening mass in matrix element, Eq.~(\ref{2.9}) takes the following form,
\begin{equation}\label{2.10}
\mid\mathcal{\bar{M}}_{HQ,q}\mid^2={\mid\mathcal{M}_{HQ,q}\mid^2}+{\mid\mathcal{M}_{HQ,q}\mid^2}^{~(1)},  
\end{equation}
with
\begin{align}\label{2.11}
{\mid\mathcal{M}_{HQ,q}\mid^2}^{~(1)}=&~512N_f\pi^2\alpha_s^2~\delta\mu^2\nonumber\\&\times\dfrac{(m_c^2-s)^2+(m_c^2-u)^2+2m_c^2t}{(t-\mu^2)^3}. 
\end{align}
Here, $\delta\mu^2$ denotes the bulk-viscous corrections to the screening mass in the QGP medium.
The bulk-viscous correction to the screening mass can be explicitly calculated from the Eq.~(\ref{1.25}) by employing the Eq.~(\ref{2.1}). Defining ${\mid\mathcal{M}_{HQ,q}\mid^2}^{~(1)}={\mid\mathcal{M}_{2}\mid^2}\delta\mu^2$ and following the same prescriptions as earlier, we have
\begin{align}\label{2.12}
\langle\langle F(p^{'})\rangle\rangle^{\text{bulk(2)}}=\Pi  B_X(X)\dfrac{1}{ 512\pi^4\gamma_c}\dfrac{1}{E_p} \Lambda_3(p, T),
\end{align}
where 
\begin{align}\label{2.13}
\Lambda&_3=\dfrac{2\alpha_s}{\pi T}\int_0^{\infty}{\dfrac{q^2}{E_q}dq}\int_{-1}^{1}{d\cos{\chi}} \dfrac{\sqrt{(s+m_c^2-m^2_{q})^2-4sm_c^2}}{s} \nonumber\\
& \times f^0_{q}(E_q)\int_{-1}^{1}{d\cos{\theta_{cm}}}\mid\mathcal{M}_{2}\mid^2 \int_{0}^{2\pi}{d\phi_{cm}e^{\beta E_{q^{'}}}f_{q}(E_{q^{'}})}\nonumber\\
&\times F(p^{'})\int_0^{\infty}{r^2~dr}\Bigg[2N_{c/f} B_M(R,T)\Big(1\pm 2{f}^0_{g/q}(E_r)\Big)\Bigg].
\end{align}
The net bulk-viscous correction to the quark contribution to the HQ drag and diffusion can be described from Eq.~(\ref{2.6}) and Eq.~(\ref{2.13}). Similarly, we incorporate the effect of bulk corrections to the screening mass in the HQ-gluon processes. In general, these corrections due to the screening mass to the HQ transport coefficients are higher order in $\alpha_s$.  Now, from Eqs.~(\ref{1.610})-(\ref{1.612}), we can define the viscous corrections to the HQ transport coefficients in the medium.
%%%%%%%%%%%%%%%%%%%%%%%%%%%%%%%%%%%%%%%%%%%%%%%%%%%%%%%%%%%%%%%%%%%
\section{Results and Discussions}
%%%%%%%%%%%%%%%%%%%%%%%%%%%%%%%%%%%%%%%%%%%%%%%%%%%%%%%%%%%%%%%%%%%
\subsection{HQ transport coefficients in the evolving QGP}
%%%%%%%%%%%%%%%%%%%%%%%%%%%%%%%%%%%%%%%%%%%%%%%%%%%%%%%%%%%%%%%%%%%
\begin{figure}
    \includegraphics[width=0.5\textwidth]{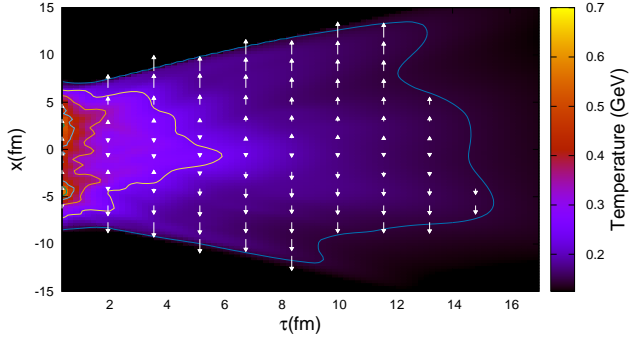}
    \caption{ Temperature evolution of QGP along the $y = 0$ axis at midrapidity. White arrows denote the size and direction of velocity fields. Curved lines indicate constant temperature contours.}
    \label{fig1}
\end{figure}

We initiate the discussion with the space-time evolution of the temperature in Pb+Pb collision at $2.76$ TeV, using the viscous hydrodynamical model-MUSIC. For this study, we have used one event from the $0-5\%$ centrality class. We use the lattice QCD based EoS from the hotQCD collaboration~\cite{Bazavov:2014pvz, Moreland:2015dvc}. Viscous effects, more specifically, terms up to the second-order gradient expansion, are incorporated in the hydrodynamical evolution. We choose $y = 0$, midrapidity to illustrate the result of our calculations. Fig. \ref{fig1} is the temperature evolution profile of our system where vectors denote the size and direction of velocity fields. As expected, the flow is larger towards the edge of the system and grows with time.
\begin{figure}
    \centering
    \includegraphics[width=0.5\textwidth]{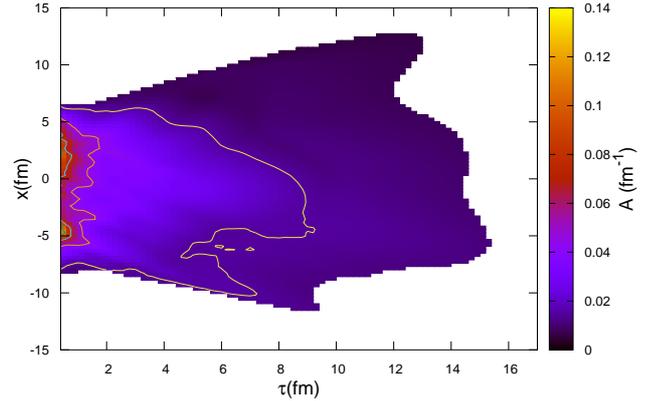}
    \caption{Drag coefficient of a charm quark with momentum $p = 5$ GeV at different space-time points. Curved lines indicate constant $A$ contours.}
    \label{fig2}
\end{figure}
\begin{figure}
    \centering
    \includegraphics[width=0.5\textwidth]{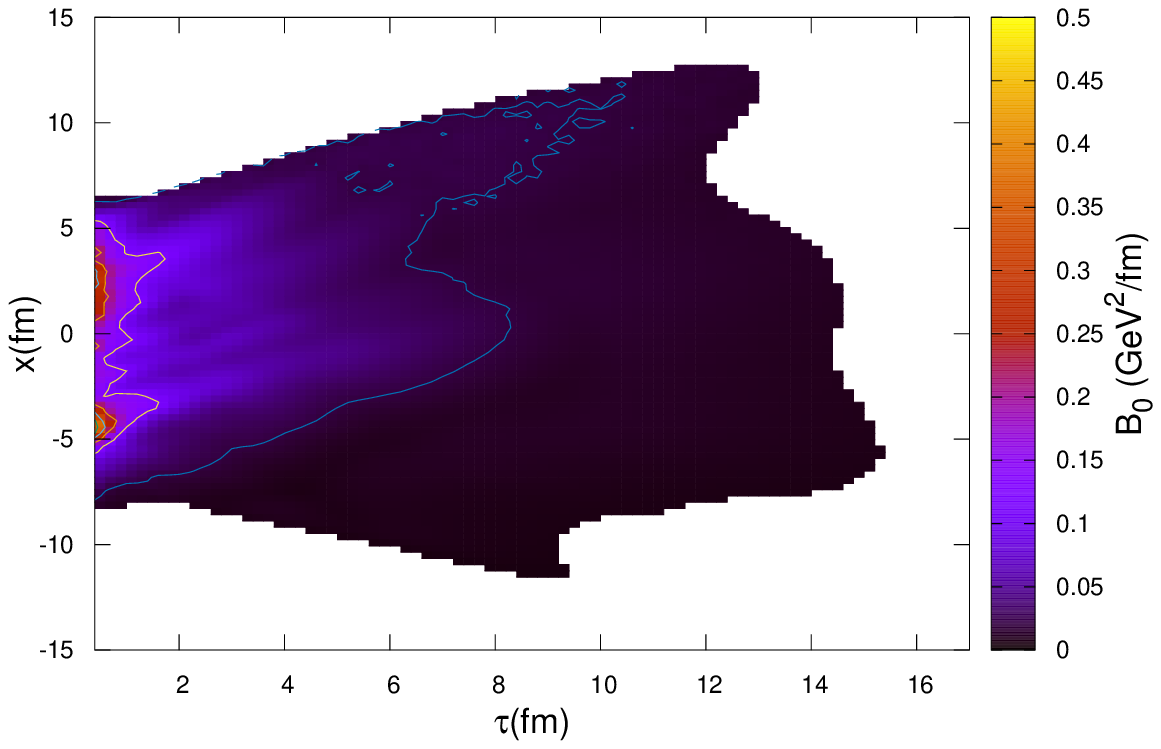}
    \includegraphics[width=0.5\textwidth]{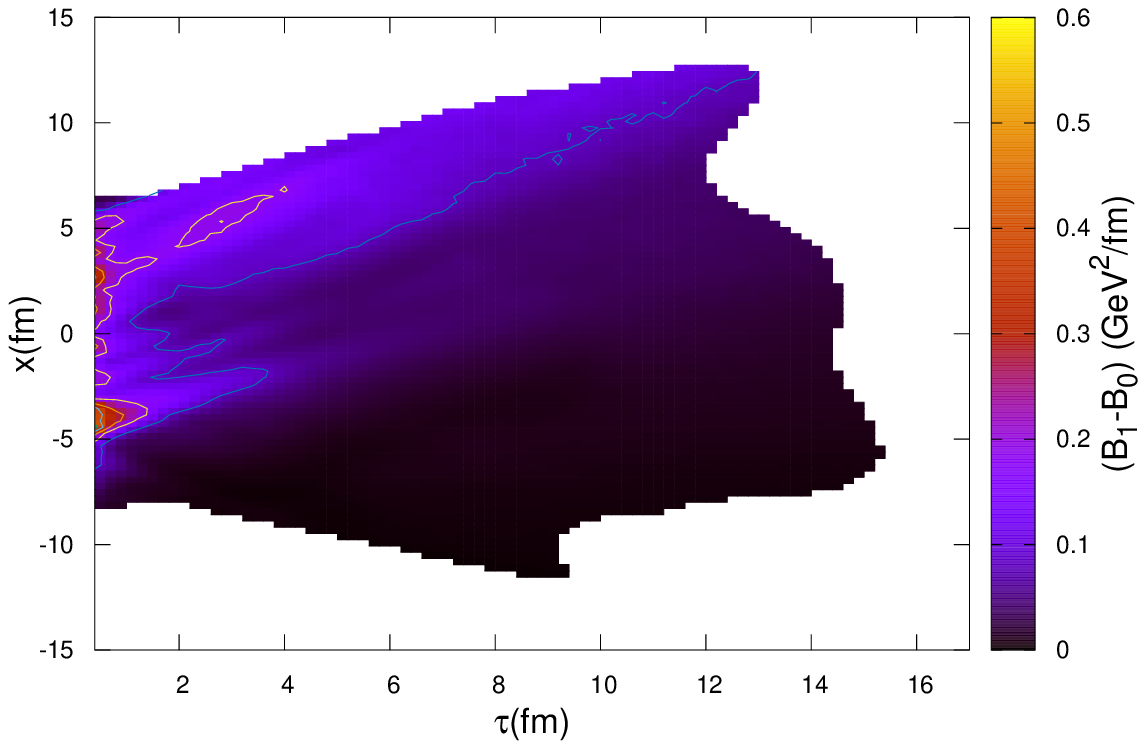}
    \caption{Diffusion coefficients $B_{0}$ (top) and $B_{1}-B_{0}$ (bottom) of a charm quark with momentum $p = 5$ GeV at different space-time points. Curved lines are constant value contours of plotted quantities.}
    \label{fig3}
\end{figure}
For the quantitative estimation of HQ drag and diffusion coefficients, we consider mass of the charm quark as $m_c=1.5$ GeV with effective number of degrees of freedom $N_f=2.5$ and the coupling constant $\alpha_s=0.3$.
The drag coefficient of the HQ with a given momentum $p=5$ GeV, at any space-time $(\tau, x)$ is shown in Fig.~\ref{fig2}. This is the drag coefficient if a charm quark with $p^{\mu} = (\sqrt{m_{c}^{2}+p^{2}}, p, 0, 0)$ with $p=5$ GeV is present at that space-time point. We observe that the drag coefficient drops as the QGP expands in space-time. This implies that the QGP offers less resistance to the HQ motion at low temperature regimes. The HQ experience more random forces in the early stage of the evolution of the QGP as compared to its equilibrated stage and hence the motion of HQ becomes more random in the medium.
This observation is qualitatively consistent with the result of Ref.~\cite{GolamMustafa:1997id}. The diffusion coefficients of the HQ with momentum $p=5$ GeV in the expanding medium is plotted in Fig.~\ref{fig3} as a function of space and time. Similar to the drag coefficient, the momentum diffusion of the HQ goes down in the low temperature regime. Further, we observe that the diffusion is larger when charm quark is moving in the same direction as the background fluid, whereas the drag is larger for charm quark moving opposite to fluid. Clearly, the details of the dynamics will play an important role.

%%%%%%%%%%%%%%%%%%%%%%%%%%%%%%%%%%%%%%%%%%%%%%%%%%%%%%%%%%%%%%%%%%%
\subsection{Effect of shear and bulk-viscous corrections to drag and diffusion}
%%%%%%%%%%%%%%%%%%%%%%%%%%%%%%%%%%%%%%%%%%%%%%%%%%%%%%%%%%%%%%%%%%%%
\begin{figure}
    \centering
    \includegraphics[width=0.5\textwidth]{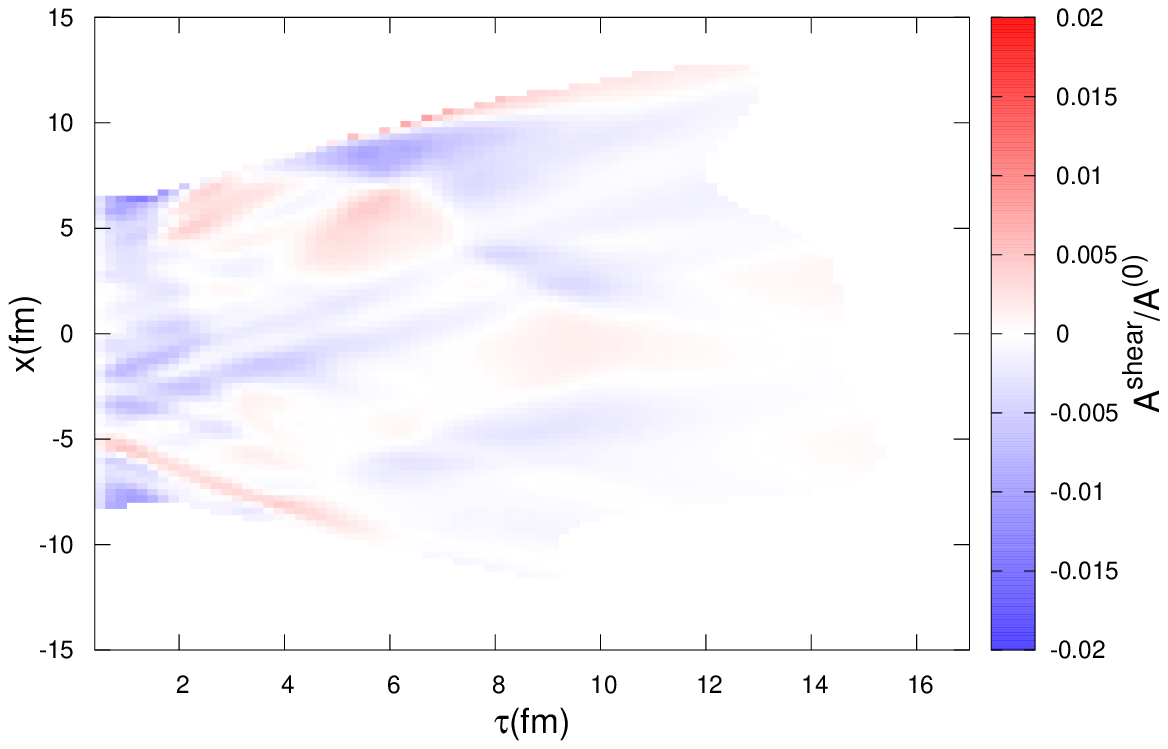}
    \includegraphics[width=0.5\textwidth]{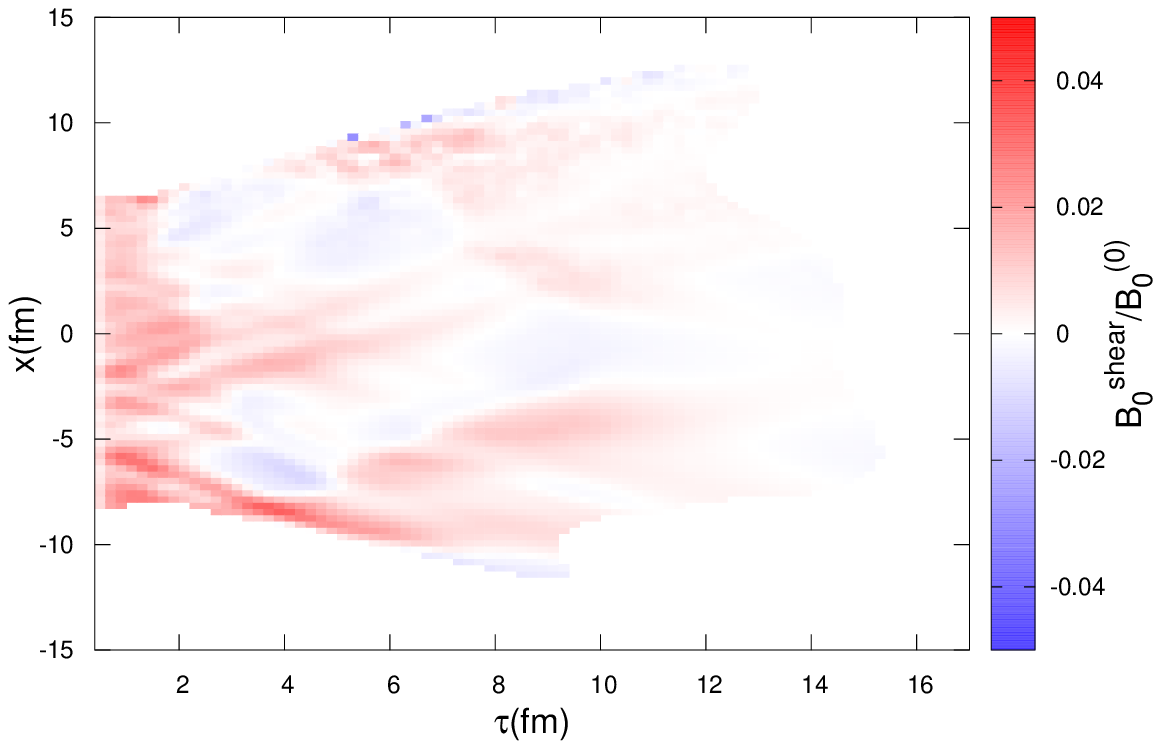}
    \includegraphics[width=0.5\textwidth]{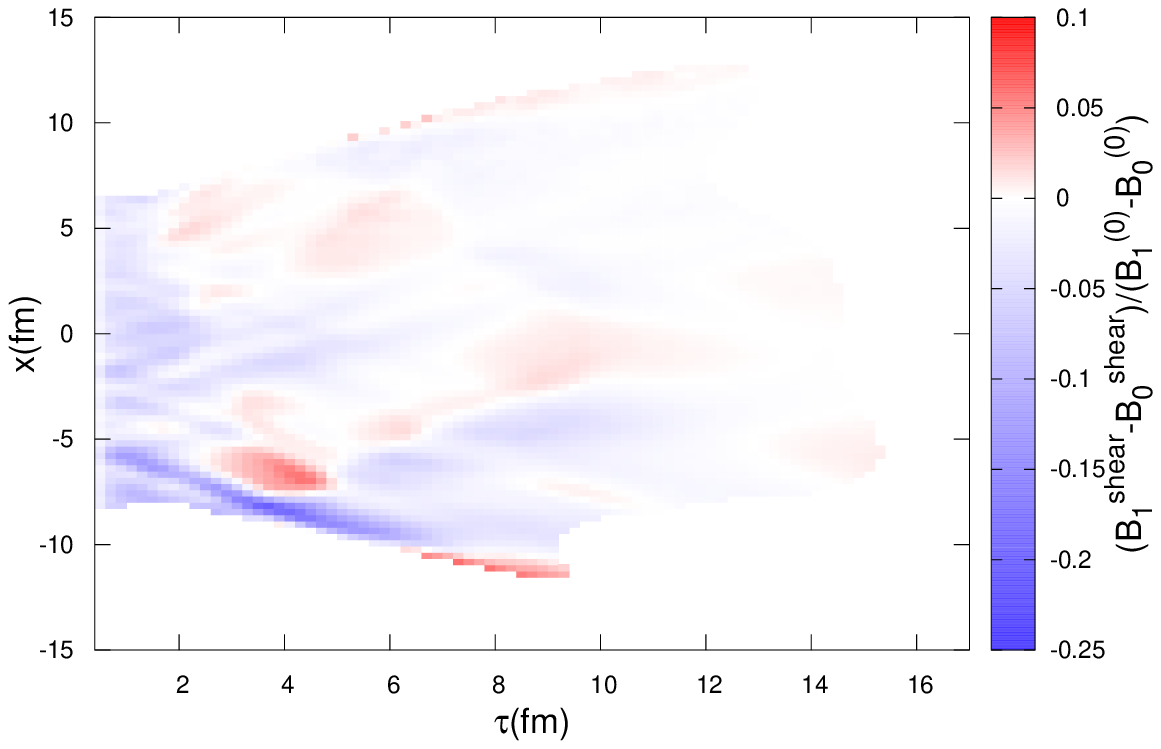}
    \caption{Ratio of shear correction to equilibrium value for drag coefficient $A$ (top), $B_{0}$ (middle) and $B_{1}-B_{0}$ (bottom).}
    \label{fig4}
\end{figure}
\begin{figure}
    \centering
    \includegraphics[width=0.5\textwidth]{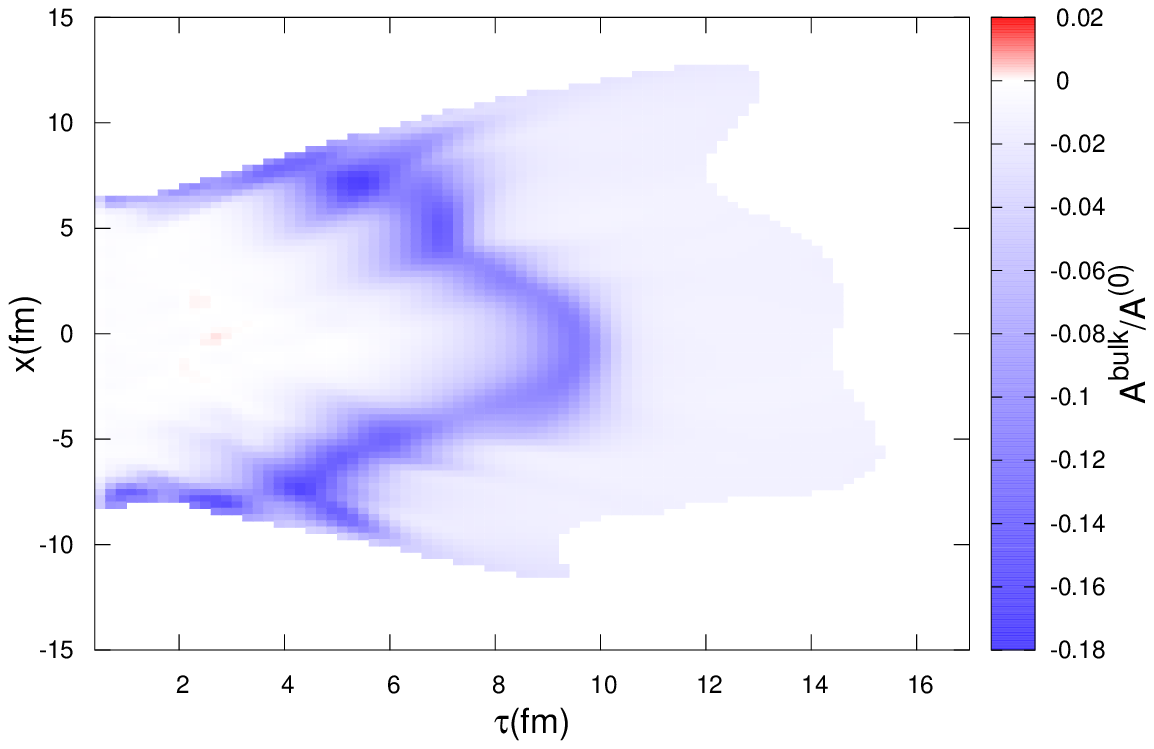}
    \includegraphics[width=0.5\textwidth]{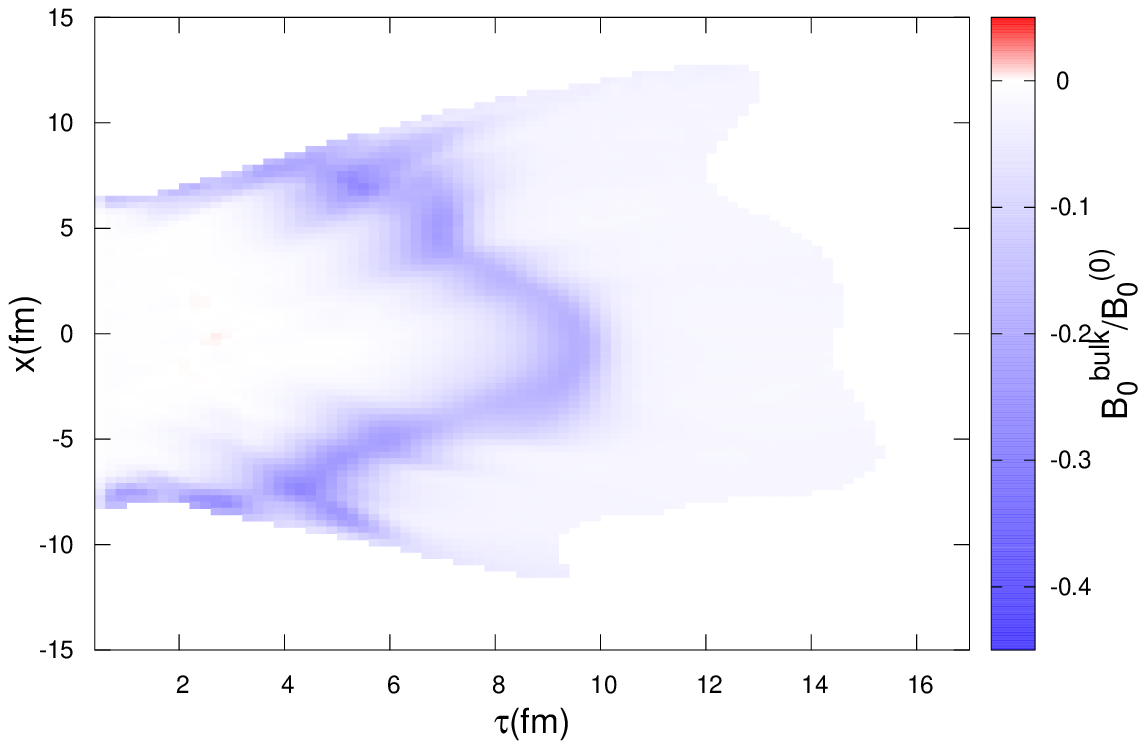}
    \includegraphics[width=0.5\textwidth]{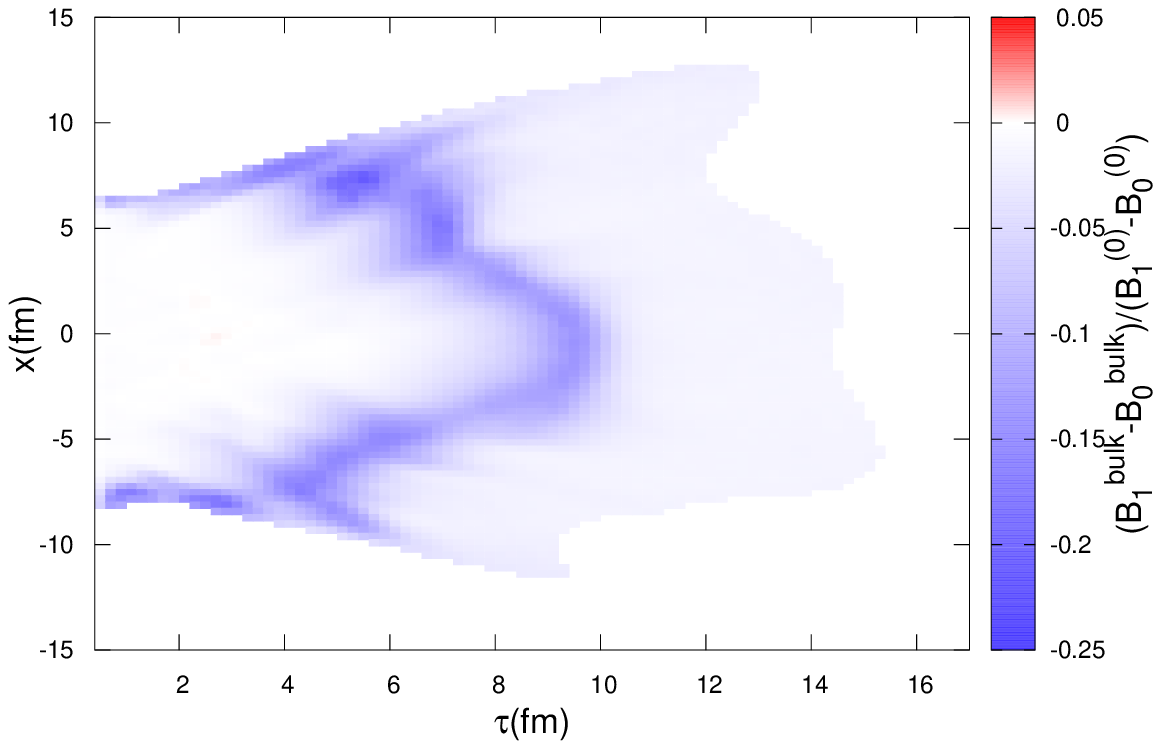}
    \caption{Ratio of bulk correction to equilibrium value for drag coefficient $A$ (top), $B_{0}$ (middle) and $B_{1}-B_{0}$ (bottom).}
    \label{fig5}
\end{figure}
We have incorporated the shear and bulk-viscous corrections through the momentum distribution function and screening mass of the QGP. The HQ drag and momentum diffusion coefficients are sensitive to the non-zero $\eta/s$, $i.e$, variation up to $10\%$ for $\eta/s=0.13$. The shear-viscous effects to the HQ drag and diffusion coefficients are studied by estimating $\frac{A^{\text{shear}}}{A^{(0)}}$, $\frac{B_0^{\text{shear}}}{B_0^{(0)}}$ and $\frac{B_1^{\text{shear}}-B_0^{\text{shear}}}{B_1^{(0)}-B_0^{(0)}}$ as shown in Fig.~\ref{fig4}. The terms $B_1^{\text{shear}}$ and $B_0^{\text{shear}}$ respectively define the first order shear-viscous correction to the longitudinal and transverse diffusion coefficients of the HQ, whereas $B_1^{(0)}$ and $B_0^{(0)}$ denote the corresponding equilibrium values. The inclusion of shear viscosity quantitatively affects the HQ transport coefficient and the effect is more pronounced for the momentum diffusion of HQs in the QGP medium. We observe that the shear-viscous effects to the drag and diffusion coefficients are negligible in the very later stage of the QGP evolution. The bulk-viscous correction to the HQ transport coefficients is depicted in Fig.~\ref{fig5}. The correction is largest when $\zeta/s$ is large. The inclusion of bulk-viscous pressure considerably modifies the HQ drag and diffusion coefficients, up to $30\%$, in the QGP medium. This observation of the significance of bulk viscosity is consistent with the recent study~\cite{Ryu:2015vwa} that highlights the large effect of the temperature dependent $\zeta/s$ in the hadronic observables in the heavy-ion collisions.

\begin{figure}[h]
    %\centering
    \includegraphics[width=0.4\textwidth]{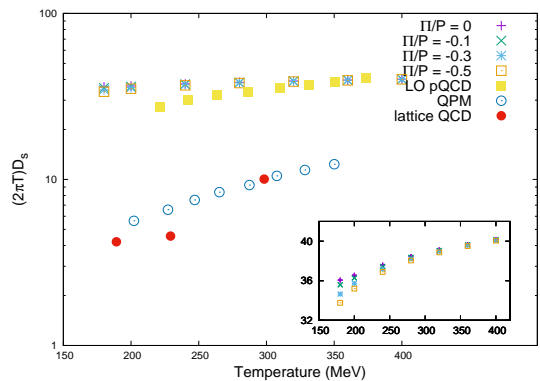}
    \caption{Spatial diffusion coefficient at the limit $p\rightarrow 0$ as a function of temperature and comparison of the results with pQCD~\cite{vanHees:2004gq}, lattice~\cite{Banerjee:2011ra}, and QPM~\cite{Scardina:2017ipo} estimations.}
    \label{fig5.1}
\end{figure}
The spatial diffusion coefficient $D_s$ is defined in the limit $p\rightarrow 0$ from the fluctuation-dissipation theorem with the form $D_s=\frac{T}{m_cA(p\rightarrow 0,T)}$. The temperature behavior of $D_s$ in the viscous QGP is depicted in Fig.~\ref{fig5.1}. In the current analysis, we are only focusing on the elastic $2\rightarrow 2$ scattering (perturbative interactions) of HQs with thermal particles in the medium via $s, t, u$ channels, and the interferences terms. Our results are consistent with those of leading order pQCD estimates in Ref.~\cite{vanHees:2004gq}. However, the $D_s$ consists of two parts, soft component and pQCD part, in which soft component accounts for the non-perturbative effects~\cite{Cao:2018ews}. The non-perturbative effects to the $D_s$ can be estimated from lattice QCD~\cite{Banerjee:2011ra} and quasiparticle model (QPM)~\cite{Scardina:2017ipo} results. The QPM incorporates the non-perturbative dynamics with a temperature-dependent background field, bag constant, and with the temperature-dependent quasiparticle mass. Note that non-perturbative effects, along with the radiation of color charges, need to be considered in the estimation of HQ observables such as nuclear suppression factor, flow coefficients, etc. This can be done by solving the Fokker-Plank equation stochastically by employing Langevin simulations, and we intend to explore this aspect in the near future. The current focus lies in the study of viscous effects to $D_s$ considering the perturbative interactions, and we observe that the viscous effects are more prominent in the temperature regime near to the transition temperature.  
%%%%%%%%%%%%%%%%%%%%%%%%%%%%%%%%%%%%%%%%%%%%%%%%%%%%%%%%%%%%%%%%%%%
\subsection{HQ energy loss in the expanding viscous medium}
%%%%%%%%%%%%%%%%%%%%%%%%%%%%%%%%%%%%%%%%%%%%%%%%%%%%%%%%%%%%%%%%%%%
\begin{figure}[h]
    \centering
    \includegraphics[width=0.51\textwidth]{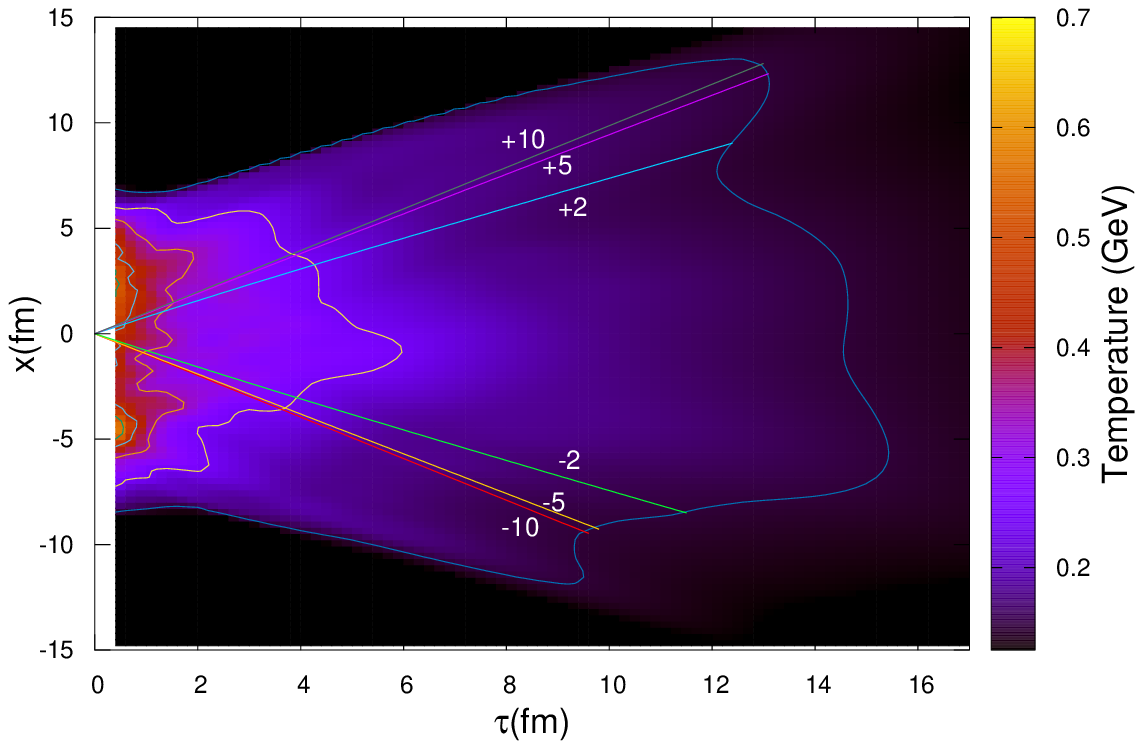}
    \includegraphics[width=0.425\textwidth]{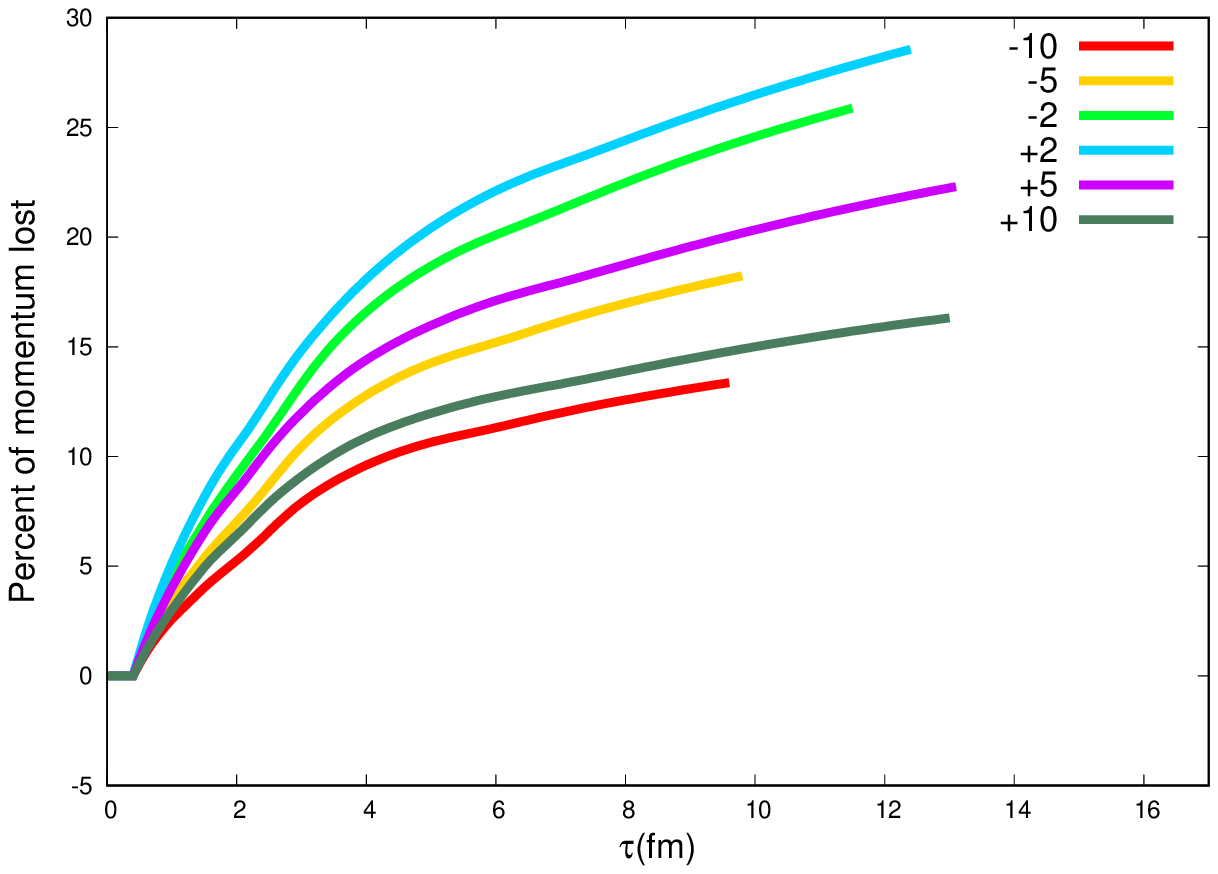}
    \caption{Different trajectories for different initial charm quark momentum (top). The charm quark momentum loss with proper time in the viscous medium at the LHC for these trajectories (bottom). The color of trajectories in the top plot corresponds to the color of momentum curves in the bottom plot.}
     \label{fig6}
\end{figure}
\begin{figure}[h]
    %\centering
    \includegraphics[width=0.425\textwidth]{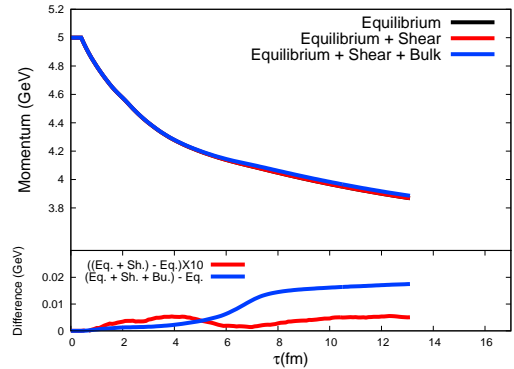}
    \caption{Charm quark momentum evolution in the QGP medium (top panel). Different colors correspond to different evolution runs with and without the inclusion of viscous corrections. The initial momentum is taken as $p=5$ GeV. Equilibrium and equilibrium + shear curves almost overlap as the effect of shear correction on energy loss is negligible. Difference in charm quark momentum between evolution runs with and without viscous corrections (bottom panel).}
    \label{fig7}
\end{figure}
HQs execute Brownian motion in the QGP medium and may lose energy by elastic collisions with quarks and gluons. 
The drag force which accounts for the resistance to the HQ motion, leads to its energy loss in the QGP medium. The differential collisional energy loss of the HQ in the QGP is related to the drag coefficient as,
\begin{equation}\label{3.1}
-\dfrac{dE}{dL}=A(p^2, T)p,
\end{equation}
where $dL$ is the length travelled by the HQ in the medium in the direction of $x$-axis within the time interval $d\tau$. A comparative study of the energy loss of the HQ from the drag force with the results of Ref.~\cite{Braaten:1991we} with hard and soft collision process is done in Ref.~\cite{GolamMustafa:1997id} and the observation confirms that the result from Eq.~(\ref{3.1}) is consistent with that of Ref.~\cite{Braaten:1991we}. To quantify the energy loss, we choose different initial momenta for the charm quark while propagating in the viscous QGP medium. Different trajectories of motion of charm quark for different initial momentum is depicted in Fig.~\ref{fig6} (top). The energy loss of the charm quark can be demonstrated by analyzing its momentum evolution in the QGP medium. The percentage of charm quark momentum loss for different trajectories (with different initial momenta) is demonstrated in Fig.~\ref{fig6} (bottom). It is observed that the charm quark loses up to $10\%-30\%$ of the initial momentum while propagating through the QGP for the duration with time interval up to $\tau_{\text{max}}=14$ fm due to the collisions with thermal particles in the medium. It is important to note that the momentum loss with proper time depends on the value of initial momentum as the drag coefficient decreases with the charm quark momentum in the medium. The momentum dependence of the HQ transport coefficients is well investigated in Refs.~\cite{Svetitsky:1987gq,GolamMustafa:1997id}. The charm quark with initial momentum $p=2$ GeV losses up to $30\%$ of its momentum while propagating in the viscous QGP whereas charm quark with $p=5$ GeV has $20\%$ of momentum loss in the QGP evolution. The viscous effects to the momentum evolution of charm quark, with initial momentum $p=5$ GeV, is plotted in Fig.~\ref{fig7}.
The viscous corrections have small effects on the momentum evolution of HQ in the medium.

%%%%%%%%%%%%%%%%%%%%%%%%%%%%%%%%%%%%%%%%%%%%%%%%%%%%%%%%%%%%%%%%%%%%
\section{Conclusion and outlook}
%%%%%%%%%%%%%%%%%%%%%%%%%%%%%%%%%%%%%%%%%%%%%%%%%%%%%%%%%%%%%%%%%%%%
In this article, we have studied the HQ dynamics in the expanding QGP medium using a realistic $3+1$D hydrodynamical modelling-MUSIC. The model describes the QGP expansion by considering the second-order evolution equations for shear tensor and bulk-viscous pressure, along with the realistic initial conditions and lattice EoS. We have described the HQ transport within the Fokker-Planck dynamics. We observe that the HQ drag and momentum diffusion coefficients drop in the later stage of the evolution of the medium. We have conducted a systematic analysis in the shear and bulk-viscous corrections to the HQ transport coefficients. The viscous corrections are incorporated through the quark and gluon phase-space  distribution functions and through the HQ-thermal particle scattering matrix element via screening mass in the analysis. The coefficients of drag and momentum diffusion of the HQ in the viscous QGP are estimated and compared to the HQ coefficients obtained in a fully thermalized medium.

Results showed that the effects of shear and bulk-viscous dynamics to the drag and diffusion are non-negligible and the variation ranges from to $0\%-30\%$ for different temperature regimes. These viscous corrections are essential to maintain consistency in the theoretical description of HQ dynamics in the QGP medium which is away from the equilibrium.
Further, we have computed the collisional energy loss of charm quark in the expanding medium at the LHC. The HQ drag force accounts for the energy loss due to the charm quark collisions with thermal particles. The energy loss of the HQ is reflected in the evolution of HQ momentum in the viscous QGP medium. We observe that the energy loss is sensitive to the initial charm quark momentum. In addition, we have investigated the effects of shear and bulk viscosities to the charm quark momentum evolution. The viscous effects are seen to have weaker dependence on the momentum evolution of the charm quark in the QGP, especially in the initial stages of the collision. A similar analysis will hold for bottom quarks, and the effects will be less pronounced because of their larger mass. The current analysis is important for the understanding of dilepton signals stemming from the decay of open charm and bottom mesons. In particular, the energy loss of the charm (bottom) quark causes a reduction in the number of high invariant mass dileptons from the decay of open charm (bottom) mesons.

The analysis presented in the article is the first step towards the investigation of the phenomenological implications of the HQ propagation in the viscous expanding medium with $3+1$D relativistic hydrodynamics. The viscous corrections to HQ transport coefficients determined in this work could affect the experimental signals such as nuclear suppression factor, elliptic flow, etc. The hydrodynamic description of the $p_T$ spectra and flow of heavy baryons could be modified by incorporating the realistic temperature dependence. We intend to work on these interesting aspects in the near future. Investigating the radiative energy loss that is almost the same order of collisional energy loss at high energy scales of HQ ($6$ GeV$-10$ GeV), and the effects of electromagnetic fields on HQ transport while including the non-equilibrium corrections are other interesting directions to follow. 
%%%%%%%%%%%%%%%%%%%%%%%%%%%%%%%%%%%%%%%%%%%%%%%%%%%%%%%%%%%%%%%%%%%
\section*{ACKNOWLEDGMENTS}
%%%%%%%%%%%%%%%%%%%%%%%%%%%%%%%%%%%%%%%%%%%%%%%%%%%%%%%%%%%%%%%%%%%%
We are thankful to Sigtryggur Hauksson, Scott McDonald, and Shuzhe Shi for useful discussions and suggestions. M.K. acknowledges the hospitality of McGill University, and the Indian Institute of Technology, Gandhinagar (IIT GN) for the Overseas Research Experience Fellowship to visit McGill University. This work was funded in part by the Natural Sciences and Engineering Research Council of Canada, by SERB for the Early Career Research Award (ECRA/2016), and by the DST, Govt. of India for INSPIRE-Faculty Fellowship (IFA-13/PH-55).
%%%%%%%%%%%%%%%%%%%%%%%%%%%%%%%%%%%%%%%%%%%%%%%%%%%%%%%%%
{}
\end{document}